\documentclass[prd,10pt,aps,twocolumn,showpacs,natbib]{revtex4}

\usepackage{times} 
\usepackage{graphicx,color}
\usepackage{longtable}
\usepackage{amssymb}

\mathchardef\bigtilde="0365

\begin{document}

\title{Delta expansion and Wilson fermion in the Gross-Neveu model: \\ Compatibility with linear divergence and continuum limit from inverse-mass expansion}
\author{Hirofumi Yamada}\email{yamada.hirofumi@it-chiba.ac.jp}
\affiliation{%
Division of Mathematics and Science, Chiba Institute of Technology, 
\\Shibazono 2-1-1, Narashino, Chiba 275-0023, Japan}

\date{\today}

\begin{abstract}
{We apply the $\delta$-expansion to the Gross-Neveu model in the large $N$ limit with Wilson fermion and investigate dynamical mass generation from inverse-mass expansion.   The dimensionless mass $M$ defined via the effective potential is employed as the expansion parameter of the bare coupling constant $\beta$ which is partially renormalized by the subtraction of linear divergence.  We show that $\delta$-expansion of the $1/M$ series of $\beta$ is compatible with the mass renormalization.  After the confirmation of the continuum scaling of the bare coupling without fermion doubling, we attempt to estimate dynamical mass in the continuum limit and obtain the results converging to the exact value for values of Wilson parameter $r\in (0.8,1.0)$.
}
\end{abstract}


\maketitle

\section{Introduction}
As well known, the system of real fermion fields on the lattice with translational invariance, chiral symmetry and locality has been found to contain unphysical redundancy \cite{nn}.   This no-go-theorem forbids any simple and straightforward fermion implementation into the lattice.  The eldest proposal to circumvent the problem is to break the chiral symmetry to the first order in the lattice spacing $a$ due to the lattice inventor K. G. Wilson \cite{wil}.  
Among other proposals \cite{rev}, Wilson's fermion has an advantage in the strong coupling expansion, since the Wilson fermion system allows an easy path to the expansion.  The strong coupling expansion provides a simple, systematic and powerful computational scheme to clarify rich physics out of reach of continuum perturbation theory.  Pertaining to the approach where the continuum limit is accessed by the strong coupling series under the crucial help of $\delta$-expansion \cite{yam,yam1,yam2}, we like to focus on Wilson fermion system and investigate the recovery of the asymptotic freedom and dynamical mass generation in 2D  lattice Gross-Neveu model in the large $N$ limit \cite{gn}.  

In the formulation with auxiliary field $\sigma_{x}$, the action of the lattice Gross-Neveu model reads ($\mu=1,2$)
\begin{eqnarray}
S&=&-\frac{a}{2}\sum_{x,\mu}\Big[\bar\psi_{x}(r-\gamma_{\mu})\psi_{x+\mu}+\bar\psi_{x+\mu}(r+\gamma_{\mu})\psi_{x}\Big]\nonumber\\
& &+2ar\sum_{x}\bar\psi_{x}\psi_{x}+a^2\sum_{x}\sigma_{x}\bar\psi_{x}\psi_{x}\nonumber\\
& &+\frac{Na^2}{2g^2}\sum_{x}(\sigma_{x}-\delta m)^2,
\label{action1}
\end{eqnarray}
where $\psi_{x}$ and $g$ stand for the $N$ flavor fermion on the site $x$ and bare coupling constant, respectively.  One choice of the explicit $\gamma$ matrices is
\begin{equation}
\gamma_{1}=\sigma_{2},\quad \gamma_{2}=\sigma_{1},\quad \gamma_{5}=\sigma_{3}=i\gamma_{1}\gamma_{2}
\end{equation}
where
\begin{equation}
\sigma_{1}=\Big(
\begin{array}{cc}
0 & 1 \\
1 & 0
\end{array}
\Big),\quad \sigma_{2}=\Big(
\begin{array}{cc}
0 & -i \\
i & 0
\end{array}
\Big),\quad \sigma_{3}=\Big(
\begin{array}{cc}
1 & 0 \\
0 & -1
\end{array}
\Big),
\end{equation}
and $\sigma_{k}^2=1$ $(k=1,2,3)$.   
The parameter $r$ is called Wilson parameter and is kept non-zero to avoid fermion species doubling.  $\delta m$ means the linearly divergent mass which is fixed by the one-loop computation.  
The fermion propagator in the momentum space ($-\pi/a<p_{\mu}<\pi/a$) reads
\begin{equation}
S_{F}(p)=\frac{1}{\sum_{\mu}i\gamma_{\mu}\frac{1}{a}\sin ap_{\mu}+\frac{r}{a}\sum_{\mu}(1-\cos ap_{\mu})},
\end{equation}
where $\mu$ takes the values $1,2$.  The added extra term in $S_{F}^{-1}$ due to Wilson behaves as $(1/2)ra\sum_{\mu}p_{\mu}^2$ near $p_{\mu}\sim 0$ and is negligible.  At a corner of Brillouin zone, $p=(\pi/a,\pi/a)$ for example, it behaves as $4r/a$ and grows as $a\to 0$.  In addition of the four corners, the extra term behaves near the boundary of Brillouin zone as $2r/a$ or $4r/a$ and behaves as the rest mass which goes to infinity and decouples in the continuum limit.

Due to the explicit breaking of the $\gamma_{5}$ symmetry at $r\neq 0$, the radiative correction for the self energy diverges linearly and the counter term represented by $\delta m\bar\psi\psi$ must be accounted.  Explicit calculation specifies that 
\begin{equation}
\delta m=-(2g^2/a) I
\end{equation}
where $I$ is given by
\begin{equation}
I(r)=\int _{-\pi}^{\pi}\frac{d^2 p}{(2\pi)^2}\frac{r\sum_{\mu}(1-\cos p_{\mu})}{\{r\sum_{\mu}(1-\cos p_{\mu})\}^2+\sum_{\mu}\sin^2 p_{\mu}}.
\end{equation}
By the introduction of the mass counter term, the fermion stays massless to all orders of perturbative expansion.  

Basic computational framework we take is the expansion in inverse powers of the mass suitably defined to be dimensionless with the combination of the lattice spacing $a$.  Largeness of the mass $M$ means the largeness of the lattice spacing $a$ and therefore the $1/M$ expansion is equivalent with the strong coupling expansion.  
The subjects to discuss on the present work are as follows:  First we like to show that the mass renormalization is compatible with the large mass expansion even when the $\delta$ expansion is applied.  This correct the wrong statement in ref. \cite{yam2} that the conventional truncation prescription of the $\delta$-expansion fails to remove the linear divergence.   We next confirm that the scaling of bare coupling (we do not  perform full renormalization including the coupling constant, since our approach depends on the bare quantities yet) is of the physical one without fermion doubling.  
Then, we demonstrate that the mass computation can be approximately carried out from large mass expansion, which is valid in the large lattice spacings.  By such a detailed analysis of the $\delta$-expansion to a solvable fermionic model, we will obtain one further evidence of the effectivity of $\delta$-expansion combined with the large mass expansion.

\section{Estimation of the mass from $1/M$ expansion}
\subsection{Overview and strategy}
The mass $M$ to be used in this work is defined through the effective potential $V(\sigma)$.  In the large $N$ limit, the fermion integration is of quadratic and results
\begin{eqnarray}
Va^2&=&\frac{a^2(\sigma-\delta m)^2}{2g^2}-\int_{-\pi}^{\pi}\frac{d^2p}{(2\pi)^2}\log\Big[\sum_{\mu=1,2}\sin^2 p_{\mu}\nonumber\\
& &+(\sigma a+r\sum_{\mu=1,2}(1-\cos p_{\mu}))^2\Big].
\end{eqnarray} 
The mass $m_{D}$ to be dynamically generated is given by $m_{D}=\sigma^{*}$ where $\sigma^{*}$ denotes the solution  of $dV/d\sigma=0$.   Then we define the dimensionless mass by
\begin{equation}
m_{D}a=M.
\end{equation}
This manifests that $1/M$ expansion is just an expansion effective at large lattice spacings.  The continuum limit is apparently the limit $M\to 0$.

The necessary condition of the ground state $dV/d\sigma=0$ gives the gap condition.  It reads in terms of $M(=a\sigma^{*})$ as
\begin{eqnarray}
M
&=&a\delta m+2g^2\int _{-\pi}^{\pi}\frac{d^2 p}{(2\pi)^2}\nonumber\\
& & \frac{M+r\sum_{\mu}(1-\cos p_{\mu})}{\{M+r\sum_{\mu}(1-\cos p_{\mu})\}^2+\sum_{\mu}\sin^2 p_{\mu}}\nonumber\\
&=&2g^2\Big[-I(r)+\int _{-\pi}^{\pi}\frac{d^2 p}{(2\pi)^2}\nonumber\\
& & \frac{M+r\sum_{\mu}(1-\cos p_{\mu})}{\{M+r\sum_{\mu}(1-\cos p_{\mu})\}^2+\sum_{\mu}\sin^2 p_{\mu}}\Big]
\label{gap1}
\end{eqnarray}
For a given positive value of $g^2$, there corresponds one value of $M(g^2,r)$.  For example, as $g^2\to 0$ in accord with the asymptotic freedom, $M\to 0$ as expected.  There appears, however, the upper limit of $M$ in the strong coupling limit $g^2\to \infty$ due to the subtraction of linear divergence (In region of large enough $M$, contribution of the counter term becomes dominant for any non-zero $r$).  The limit $M(\infty,r)$ is smaller for larger $r$ and larger for smaller $r$.  For instance, at $r=1$, $M(\infty,1)=0.46732772346\cdots$.  In the limit $r\to 0$, $M(\infty,r)\to \infty$.  Thus,  $1/M$ expansion covers both physical and unphysical regions. 

Now, from (\ref{gap1}), it follows that
\begin{eqnarray}
\beta&:=&\frac{1}{2g^2}\nonumber\\
&=&\int _{-\pi}^{\pi}\frac{d^2 p}{(2\pi)^2}\frac{1+r/M\sum_{\mu}(1-\cos p_{\mu})}{\{M+r\sum_{\mu}(1-\cos p_{\mu})\}^2+\sum_{\mu}\sin^2 p_{\mu}}\nonumber\\
& &-\frac{I(r)}{M}.
\label{gap2}
\end{eqnarray}
For naive fermion at $r=0$, $I(0)=0$ and (\ref{gap2}) gives $\beta=\int _{-\pi}^{\pi}\frac{d^2 p}{(2\pi)^2}\frac{1}{M^2+\sum_{\mu}\sin^2 p_{\mu}}$.  Thus, $\beta$ becomes a function in the square of $M$ rather than $M$ itself.  In the bosonic case like non-linear $\sigma$ models and Ising models, the inverse coupling constant or the inverse temperature is described in the square of the mass.   There exists a discrepancy in the suitable mass parameter between the present model at $r\neq 0$ and bosonic models.

Though unphysical region is also covered, $1/M$ expansion for $\beta$, which we denote as $\beta_{>}$, is readily obtained as
\begin{eqnarray}
\beta_{>}&=&-\frac{1}{M}I(r)\\
& &+\Big\{\frac{1}{M^2}+\frac{-2r}{M^3}+\frac{-1+5r^2}{M^4}+O(M^{-5})\Big\}.\nonumber
\label{betagap}
\end{eqnarray}
The above expansion becomes useless beyond the $r$-dependent convergence radius, and the small $a$ behavior cannot be accessed.   As we can see below, the $\delta$-expansion changes the status in a drastic manner.  

Suppose that $\beta_{>}$ is truncated at order $n$ such that $\beta_{n>}=\sum_{k=1}^{n}b_{k}/M^k$.  The result of the $\delta$-expansion is summarized by the formula
\begin{equation}
M^{-k}\to { n \choose k}t^k
\label{delta1}
\end{equation}
with the binomial coefficient
\begin{equation}
{n \choose k}=\frac{n!}{k!(n-k)!}.
\end{equation}
The $\delta$-expansion induces an order-dependent transformation from $\sum_{k=1}^{n}b_{k}/M^k$ to the truncated series in $t$, $\sum _{k=1}^{n}b_{n}{n \choose k}t^k$.  Let us use the notation $D[\beta_{n>}]$ or simply $\bar\beta_{n>}$ for the transformation of $\beta$.  Then
\begin{eqnarray}
\bar\beta_{n>}(t)&=&-I(r){n \choose 1}t+\Big\{{n \choose 2}t^2-2r{n \choose 3}t^3\nonumber\\
& &+(-1+5r^2){n \choose 4}t^4+\cdots +b_{n}{n \choose n}t^n\Big\}.
\label{largebeta}
\end{eqnarray}
Here $b_{n}$ stands for the coefficient of $\beta_{>}$ at $M^{-n}$.   Crucial advantage of $\bar\beta_{n>}$ is that it exhibits the scaling behavior within {\it its effective region at small $t$}.  The rigorous specification of the effective region is not known but actually, the plots shown in FIG. 1 exhibit the expected logarithmic continuum scaling of $\bar\beta_{n>}$.   
\begin{figure}[h]
\centering
\includegraphics[scale=0.75]{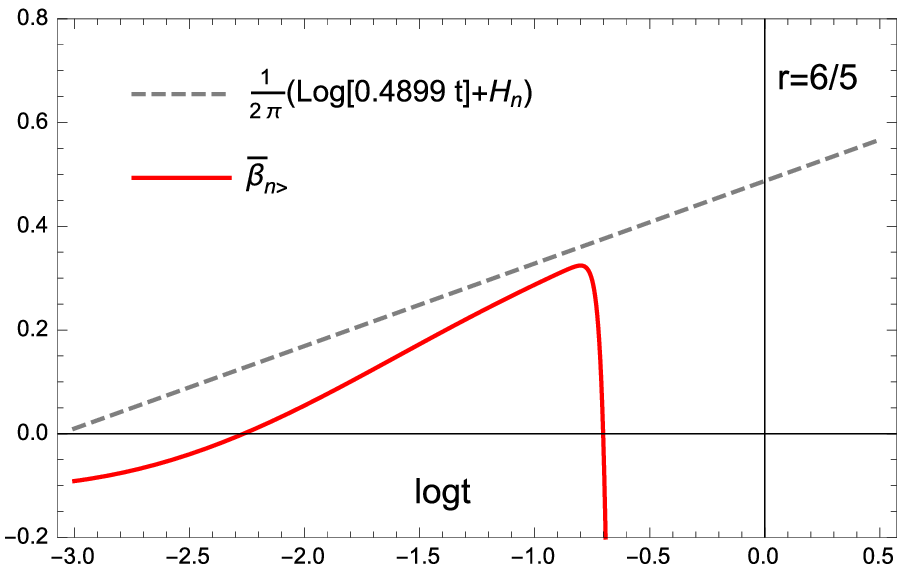}
\includegraphics[scale=0.75]{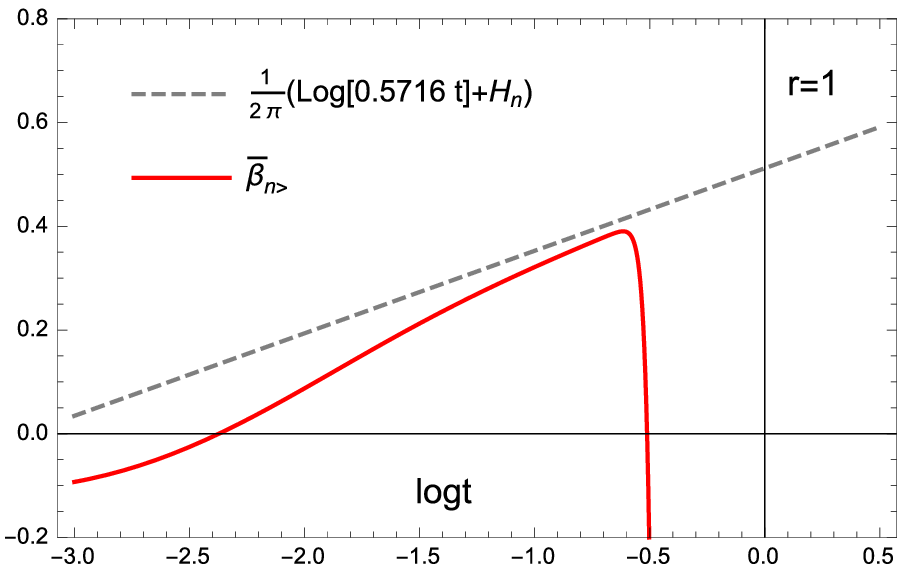}
\includegraphics[scale=0.75]{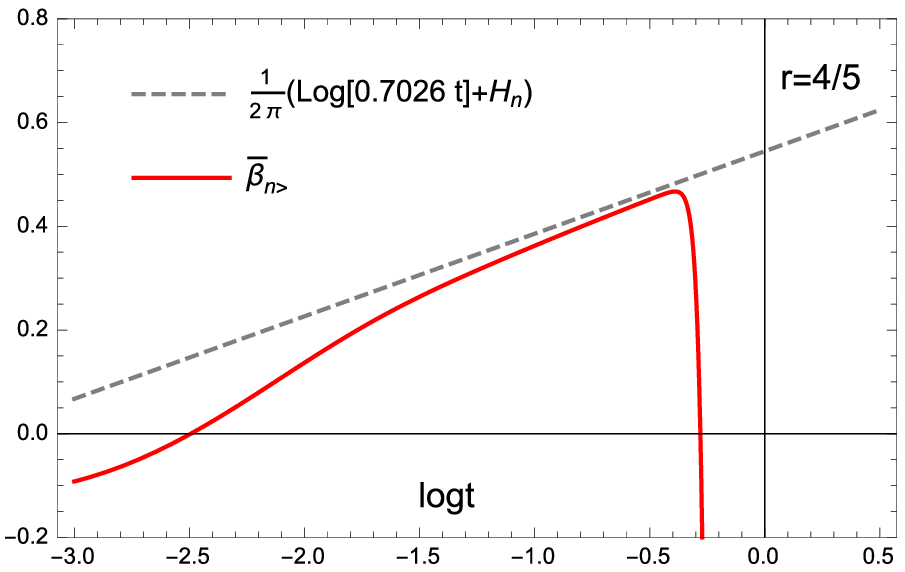}
\includegraphics[scale=0.75]{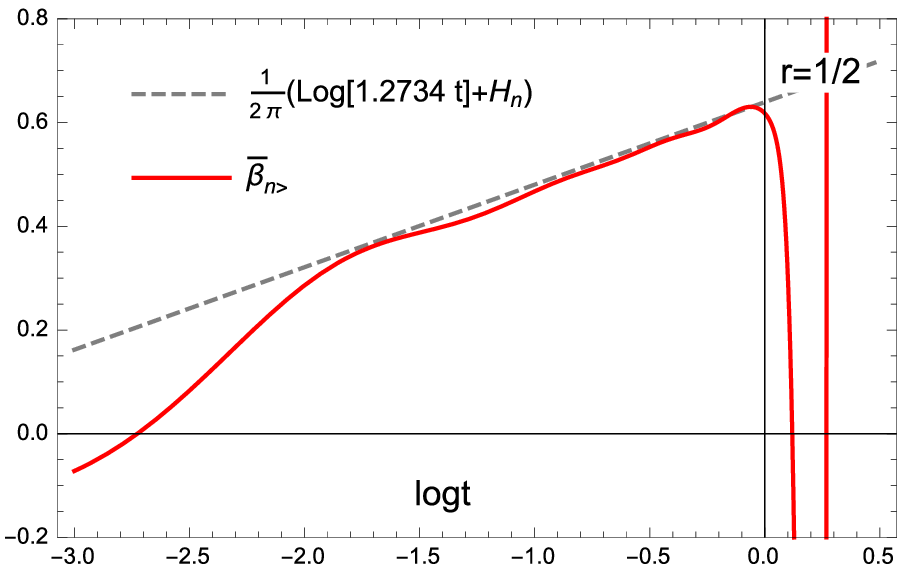}
\caption{Plots of $\bar\beta_{24>}$ and $\bar\beta_{24<}\sim 1/(2\pi)(\log (C t)+H_{24})$ at $r=6/5$ ($C=0.4899$), $r=1$ ($C=0.5716$), $r=4/5$ ($C=0.7026$) and $1/2$ ($C=1.2731$).  In the plot, we showed dashed line which indicates the continuum behavior at respective value of the Wilson parameter.  We note that the harmonic constant added to the logarithmic term comes from the $\delta$-expansion.}
\end{figure}
As $r$ decreases from $6/5$, the effective region of $\bar\beta_{n>}$ grows broader.  However, as found from the $4$th plot at $r=1/2$, the behavior of $\bar\beta_{n>}$ becomes oscillatory.  The oscillatory behavior becomes stronger for lower $r$.  We remark that the oscillatory behavior is a particular property in $\bar\beta_{n>}$, the $\delta$ expansion of the $1/M$ expansion, and not observable in the original exact function $\beta(M)$ given by (\ref{gap2}).

The oscillation shows the fluctuation about the scaling behavior and makes the estimation be too complicated.  Thus, we understand that there are preferred values of $r$.  Within the range of the preferred values, the asymptotic freedom behavior of the bare coupling is observed and the matching of the behavior of $\bar\beta_{n>}$ with $\delta$-expanded $\beta$ in the scaling region would enable us to estimate critical quantities in the continuum limit.  Though in ref. \cite{yam2}, we have wrongly stated that the cancellation of the first term in (\ref{largebeta}) (the counter term contribution from the linear divergence) with the expanded series, the rest set of (\ref{largebeta}), is incomplete, we here correct it such that the renormalization of the linear divergence is effective under the $\delta$-expansion \cite{comment}.

The most information of the scaling behavior near the continuum limit is governed by the ultraviolet structure of the model.  For example, the logarithmic behavior with the coefficient $1/(2\pi)$ is found from the perturbative expansion.  On the other hand, the nonperturbative information such as the value of the dynamical mass cannot be reached from the sole results of perturbative series.  In the Gross-Neveu model in the large $N$ limit, the only quantity of nonperturbative nature included in the bare coupling is the dynamical mass to be generated.  We like to show that the information in $\bar\beta_{n>}$ effective at small $t$ is enough to estimate the dynamical mass $m_{D}$.  To make estimation of $m_{D}$ be simpler and accurate, we use a perturbative information of the ultraviolet divergence near the continuum limit in what follows.

The behavior of the bare coupling constant is found from perturbative renormalization group that 
\begin{equation}
\beta(M)\sim \frac{1}{2\pi}\log(M/C),
\end{equation}
with unknown constant $C$.  Since the bare coupling should behave as $\beta(M)\sim \log(a\Lambda_{L})/(2\pi)$ in the $M\to 0$ limit, we find
\begin{equation}
m_{D}=C\Lambda_{L}
\end{equation}
where $\Lambda_{L}$ stands for the mass scale on the square lattice.  Thus the estimation of the constant $C$ directly gives the dynamical mass.

   The non-perturbative constant $C$ depends on the Wilson parameter and consequently $\Lambda_{L}$ too as $m_{D}$ being universal.  The non-universality of the scale $\Lambda_{L}$ is natural since it depends on the microscopic construction of the lattice model.  Analytically, $C$ is obtained by the limit, $C=\lim_{M\to 0}\exp(\log M-2\pi\beta(M))$.  By the expansion of $\beta(M)$ given by (\ref{gap2}) in the mass $M$, we then obtain
\begin{equation}
C=\exp[2\pi(c_{1}-2c_{2})],
\end{equation}
where
\begin{widetext}
\begin{eqnarray}
c_{1}&=&\lim_{M\to 0}\Big[\int_{-\pi}^{\pi}\frac{d^2 p}{(2\pi)^2}\frac{1}{(M+r\sum_{\mu}(1-\cos p_{\mu}))^2+\sum_{\mu}\sin^2 p_{\mu}}+\frac{\log M}{2\pi}\Big]\\
c_{2}&=&\int_{-\pi}^{\pi}\frac{d^2 p}{(2\pi)^2}\bigg[\frac{r\sum_{\mu}(1-\cos p_{\mu})}{(r\sum_{\mu}(1-\cos p_{\mu}))^2+\sum_{\mu}\sin^2 p_{\mu}}\bigg]^2.
\end{eqnarray}
\end{widetext}
From numerical integration, we find that $C=0.4899, 0.5716, 0.7026, 1.2731$ for $r=6/5,1,4/5,1/2$, respectively.  In experimental plots in FIG 1, we have used these values.

  In the large $N$ limit, the higher order corrections to the leading log is the lattice artifact.  The lattice artifact contains the power of $M$ and in addition the log of the mass $M$ multiplied to that.  These terms can be computed if we use the closed result (\ref{gap2}).   However, using the detailed information is not of real significance in our study.  The aim of our study is to use only accessible information in the perturbation theory and the large $M$ expansion.  Hence, we here assume that
\begin{equation}
\beta_{<}(M)=\frac{1}{2\pi}\log(M/C)+R,
\end{equation}
where $R$ denotes the lattice artifact obeying $\lim_{M\to 0}R=0$.  The subscript "$<$" means the expansion at small $M$.  For the matching of $\bar\beta_{n>}$ with the $\delta$-expanded $\beta_{<}$, $D[\beta_{<}]=\bar\beta_{n<}$, we employ the extension of the binomial coefficient by the Gamma functions,
\begin{equation}
M^{\lambda}\to {n \choose -\lambda}t^{-\lambda},
\label{g-binomial}
\end{equation}
where
\begin{equation}
{n \choose -\lambda}=\frac{\Gamma(n+1)}{\Gamma(-\lambda+1)\Gamma(n+\lambda+1)}.
\end{equation}
Here $\lambda$ denotes any real number.  Taking $\lambda$ infinitesimal in (\ref{g-binomial}), we obtain $1\to 1$ and $\log M\to -\log t-H_{n}$ where the harmonic number $H_{n}$ is given by
\begin{equation}
H_{n}=\sum_{k=1}^{n}\frac{1}{k}.
\end{equation}    
The $\delta$-expansion on $\beta_{<}(M)$ to the order $n$ thus provides the transform
\begin{equation}
\bar\beta_{n<}(t)=-\frac{1}{2\pi}\{\log(Ct)+H_{n}\}+\bar R_{n}.
\label{smallbeta}
\end{equation}
The matching of $\bar\beta_{n>}$ and $\bar\beta_{n<}$ enables us to estimate the constant $C$ which directly gives the dynamical mass.  The matching process is conveniently carried out through the use of linear differential equation (LDE) to be approximately satisfied by $\bar\beta_{n<}$.   The construction of the LDE needs, in a strict sense, the information of the lattice artifact $\bar R$.  Though the explicit expansion from the gap equation (\ref{gap2}) proves the existence of $M^{\ell}\log M$ as mentioned before, we simply forget it and proceed in the robust manner to mimic the complicated structure with the simple power like corrections, 
\begin{equation}
\bar R=c_{1} t^{-p_{1}}+c_{2} t^{-p_{1}}+\cdots,
\end{equation}
where $0<p_{1}<p_{2}<\cdots$.  
Here we do not restrict the exponent $p_{k}$ be an integer but leave to take any positive real number.  In the estimation process of $C$ via LDE, the values of exponents will be optimized to non-integer value for the best matching.

Truncation of $\bar R$ to the first order gives $\bar\beta_{n<}=-\frac{1}{2\pi}\{\log(Ct)+H_{n}\}+c_{1}t^{-p_{1}}$.  The exponent of the logarithmic term is considered as zero of double degeneracy.  Thus, the ansatz with one-parameter obeys
\begin{equation}
\Big[0+\frac{d}{d\log t}\Big]^2\Big[p_{1}+\frac{d}{d\log t}\Big]\bar\beta_{n<}=0.
\end{equation}
In the matching region where the above LDE is valid, the function $\bar\beta_{n>}$ is also  effective and approximately satisfies the same LDE as long as the order $n$ is large enough.  Hence, for large  $n$, we deal with the same LDE for $\bar\beta_{n>}$,
\begin{equation}
\Big[0+\frac{d}{d\log t}\Big]^2\Big[p_{1}+\frac{d}{d\log t}\Big]\bar\beta_{n>}= 0,
\label{lde2}
\end{equation}
and the integration over $\log t$ provides
\begin{equation}
\Big[1+p_{1}^{-1}\frac{d}{d\log t}\Big]\bar\beta_{n>}=-\frac{1}{2\pi}\{\log(Ct)+H_{n}\},
\end{equation}
and
\begin{equation}
\Big[1+p_{1}^{-1}\frac{d}{d\log t}\Big]\bar\beta_{n>}+\frac{1}{2\pi}(\log t+H_{n})=-\frac{1}{2\pi}\log C.
\label{zero_cond}
\end{equation}
To estimate $C$, we need to input values of $p_{1}$ and $t$ around which point the LDE is considered to be satisfied.  To obtain such optimal set of $(p_{1},t)$, we employ the extension of the principle of minimum sensitivity (PMS) \cite{steve,knp}.  We first demand that the estimation should be done at the point $t$ where the left-hand side of (\ref{zero_cond}) is stationary with respect to $t$.  We further demand that the reliable estimation point should be in the scaling region, shown in this case as the plateau.   Then, it is natural to employ the second derivative of the left-hand side of (\ref{zero_cond}) be zero or approximately zero at the best estimation point.   These conditions are written as
\begin{eqnarray}
\Big[1+p_{1}^{-1}\frac{d}{d\log t}\Big]\bar\beta_{n>}^{(1)}+\frac{1}{2\pi}&=&0,
\label{first_cond}\\
\Big[1+p_{1}^{-1}\frac{d}{d\log t}\Big]\bar\beta_{n>}^{(2)}&\sim&0
\label{second_cond}
\end{eqnarray}
The symbol "$\sim$" in (\ref{second_cond}) means the exact or the approximate equality (when a close point to zero exists).  Note that the second condition (\ref{second_cond}) is nothing but (\ref{lde2}).  
One can first solve (\ref{first_cond}) to give $p_{1}$ as the function of the stationary point $t$, $1/p_{1}=\rho(t)$.  Then, substituting the solution into the left-hand-side of (\ref{second_cond}), we can obtain one or a few solutions.  Among them, the optimal one is identified by the value of $t$ where all relevant functions $\bar\beta_{n>}^{(\ell)}$ for $\ell=0,1,2,3$ show expected scalings.  For example, $\bar\beta_{n>}$ at $r=1$ shows approximate scaling about $t\sim 0.5$ and optimal solution $t^{*}$ should be found around there.   In these natural criteria, we can obtain only one solution at each order.  Using the optimal solution $t^{*}$, we obtain $1/p_{1}^{*}=\rho(t^{*})$ and from (\ref{zero_cond})
\begin{equation}
C=\exp\Big[-2\pi(\bar\beta_{n>}+1/p_{1}^{*}\bar\beta_{n>}^{(1)})|_{t^{*}}
-(\log t^{*}+H_{n})\Big]
\end{equation}

It is also possible to incorporate the next order correction $t^{-p_{2}}$.  Then, LDE with which we start reads $[0+\frac{d}{d\log t}]^2[p_{2}+\frac{d}{d\log t}][p_{1}+\frac{d}{d\log t}]\bar\beta_{n<}= 0$ and
\begin{eqnarray}
& &\Big[1+p_{2}^{-1}\frac{d}{d\log t}\Big]\Big[1+p_{1}^{-1}\frac{d}{d\log t}\Big]\bar\beta_{n>}\nonumber\\
& &+\frac{1}{2\pi}(\log t+H_{n})=-\frac{1}{2\pi}\log(C).
\label{est-2}
\end{eqnarray}
The extended PMS conditions read
\begin{eqnarray}
\Big[1+p_{2}^{-1}\frac{d}{d\log t}\Big]\Big[1+p_{1}^{-1}\frac{d}{d\log t}\Big]\bar\beta_{n>}^{(1)}+\frac{1}{2\pi}&=&0,
\label{2first_cond}\\
\Big[1+p_{2}^{-1}\frac{d}{d\log t}\Big]\Big[1+p_{1}^{-1}\frac{d}{d\log t}\Big]\bar\beta_{n>}^{(2)}&=&0,
\label{2second_cond}\\
\Big[1+p_{2}^{-1}\frac{d}{d\log t}\Big]\Big[1+p_{1}^{-1}\frac{d}{d\log t}\Big]\bar\beta_{n>}^{(3)}&\sim&0.
\label{2third_cond}
\end{eqnarray}
From the first two conditions, we obtain $p_{1}^{-1}+p_{2}^{-1}=\rho(t)$ and $(p_{1}p_{2})^{-1}=\sigma(t)$ as functions of $t$ and then, from the third condition, optimal $t=t^{*}$ can be obtained within the observable scaling region.  Following the same manner as in the case of one-parameter ansatz, we then obtain $p_{1}^{*}$ and $p_{2}^{*}$ and $C$ from (\ref{est-2}).   

Next order correction $t^{-p_{3}}$ is difficult to incorporate, since the necessary higher order derivatives $\bar\beta^{(\ell)}_{n>}$ $(\ell=6, 7)$ do not show scalings even at $n=50$ which is our limit, for practical reason of computer facility.  

\subsection{Estimation at $r=1$}
We first confine ourselves with the popular choice $r=1$.  
The result of estimation up to the $2$-parameter ansatz is summarized in Table 1 and FIG.  2(a).  In FIG 2, estimation results of $p_{1}^{-1}$ are also plotted in (b).
\begin{table}
\caption{Estimation result of $C=0.5716061\cdots$ in 1- and 2-parameter ansatz at $r=1$.}
\begin{center}
\begin{tabular}{ccccc}
\hline\noalign{\smallskip}
${\small {\rm order}\,\, n}$ & {\small 20} & {\small 30} & {\small 40} & {\small 50} \\
\noalign{\smallskip}\hline\noalign{\smallskip}
{\small 1-parameter} &  {\small 0.5505666}  & {\small 0.5610658} & {\small 0.5666871} &  {\small 0.5688271}  \\
{\small 2-parameter} &   & {\small 0.5873315} & {\small 0.5708970}  & {\small 0.5709232}  \\
\noalign{\smallskip}\hline
\end{tabular}
\end{center}
\end{table}

\begin{figure}
\centering
\includegraphics[scale=0.8]{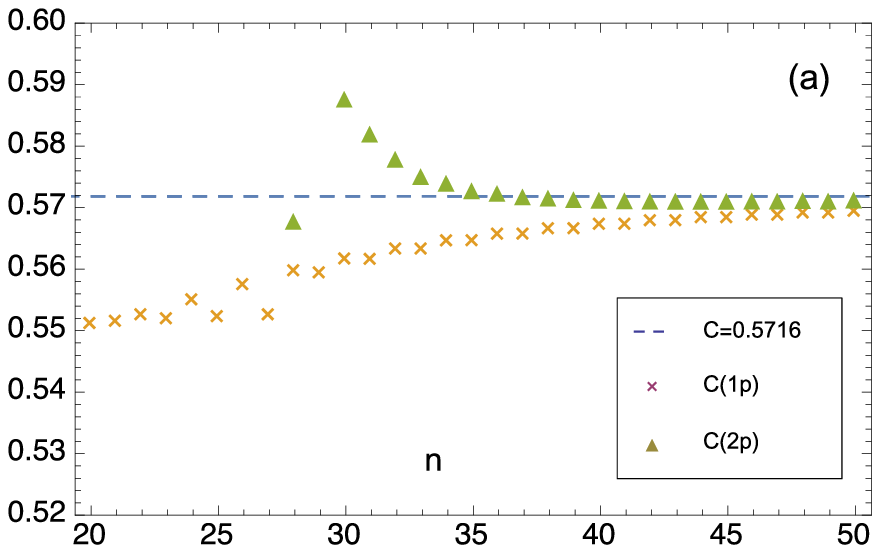}
\includegraphics[scale=0.8]{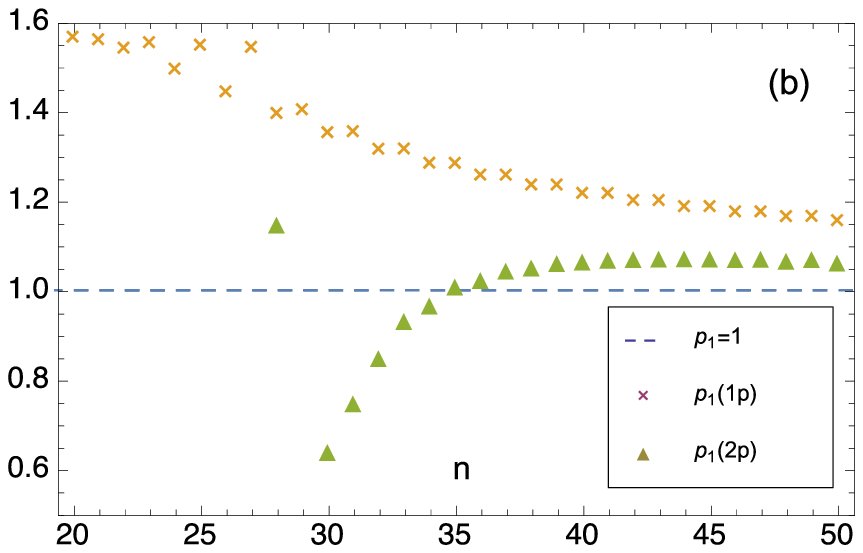}
\caption{Estimation results of $C=0.5716061\cdots$ (figure (a)) and $p_{1}=1$ (figure (b)) in 1- and 2-parameter andatz at $r=1$.  Plotted estimation is for $n=20$ to $50$.}
\end{figure}
We find that the sequence of the $C$-estimate shows tendency to the exact value $C=0.5716$.  The speed of convergence is rather slow in $1$-parameter ansatz.  The estimate in $2$-parameter ansatz yields accurate value but the onset of reliable estimation starts around order $35$th.  The result of $p_{1}$ estimate shown in FIG. 2(b) has been obtained from the work of $C$-estimation as a byproduct.  The limit of the sequence suggested is not clear yet.  However, it is roughly approaching to the value $1$, which is actually the exponent of  the leading order term in $\bar R$:  In fact, from the exact result (\ref{gap2}), $R$ is given by $R=M(const+const\log M)+O(M^2)$.  Then $\bar R=const \times(1/t)+O(t^{-2})$, giving $p_{1}=1$.

We have explored the possibility of the direct estimation of $p_{1}$ through $\bar\beta^{(3)}/\bar\beta^{(2)}$ showing scaling $\sim -p_{1}$.  However, we failed because the ratio function shows large oscillation.

\subsection{Estimation at $r\neq 1$}
Now, we discuss on the estimation work for $r\neq 1$.   One might consider that the case $r=1/2$ may provide better values since the function $\bar\beta_{n>}$ is closer to $\bar\beta_{n<}$, as seen in the last plot in FIG 1.  However, $\bar\beta_{n>}$ slightly oscillates at $r=1/2$ and the derivatives would show oscillations with larger amplitudes.  Actually, by explicit plots of $\bar\beta_{n>}^{(\ell)}$ $(\ell=1,2)$, we find that LDE approach does not work well due to the disturbing oscillation.  Since the incorporation of the derivatives is crucial for accurate estimation, this throws us a severe problem.    From the plots of $\bar\beta_{n>}$ and the derivatives, we arrive at the following observation.

When $r$ is larger than $1$, the oscillation is absent but the effective range of $\bar\beta_{n>}$ is narrow, giving less accurate estimate of $C$.  
When $r$ is less than and close to $1$, we observe weak oscillatory property in $\bar\beta_{n>}$ and the derivatives at low orders.  Though this makes the estimation slightly complicated but the confirmation of the scaling region is possible.   Around $r\sim 0.8$ or slightly larger region, the scaling behavior in $\bar\beta^{(\ell)}$ is roughly visible in low order derivatives and the estimation protocol by extended PMS is permitted.  The value $r=0.8$ is best in the sample three values.  We confirmed that $r$ smaller than $0.8$ makes the estimation worse due to the growing oscillation.   This is the reason that the effective range of Wilson parameter is approximately found to be $(0.8,1.0)$.  The result of estimation is depicted in FIG 3.

We report the result of $p_{1}$ estimation by $p_{1}^{*}$.  The result is shown in FIG 4.  We find that, in the 1-parameter ansatz, the value $r=0.8$ produces best estimation in the typical three values $r=0.8,0.9$ and $1.0$.  At $2$-parameter case, however, the ansatz $r=0.8$ produces somewhat unstable and oscillatory sequence.   This may be the sign that the smaller $r$ is not adequate for the estimation using higher order derivatives.

\begin{figure}
\centering
\includegraphics[scale=0.8]{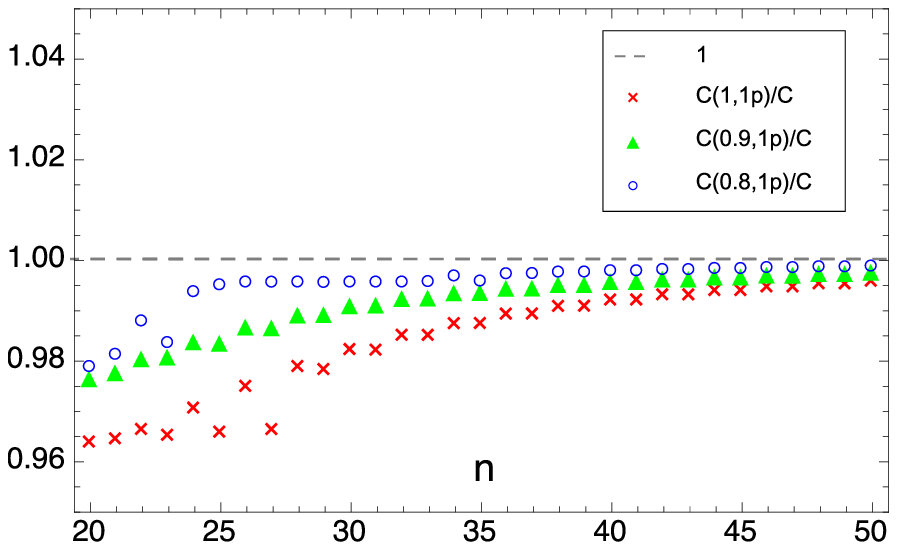}
\includegraphics[scale=0.8]{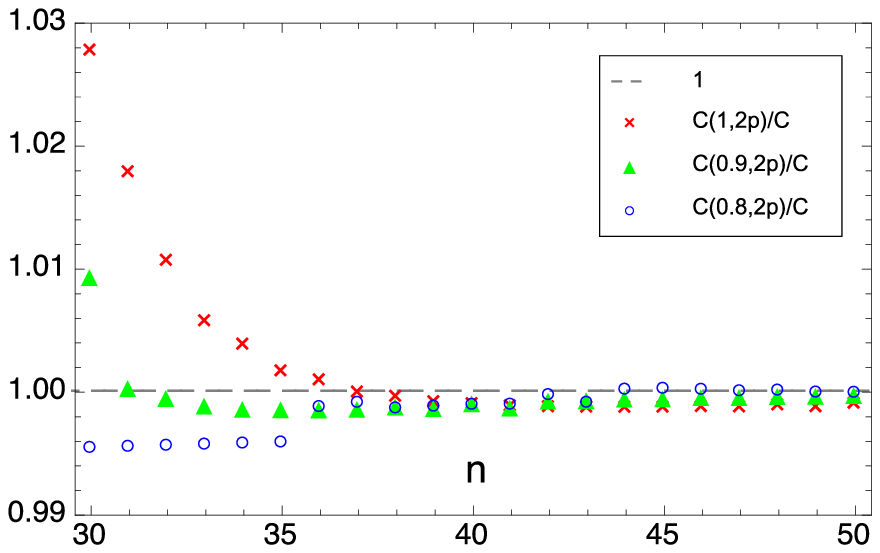}
\caption{Estimation results of $C$ for $r=1,0.9,0.8$ at $1$- and $2$-parameter ansatze (respectively plotted in the upper and lower).  Here $C=0.5716\cdots$, $C=0.62797\cdots$ and $C=0.7026\cdots$ for $r=1, 0.9, 0.8$ respectively.  The plots show the ratio of estimate to the exact values.}
\end{figure}

\begin{figure}
\centering
\includegraphics[scale=0.8]{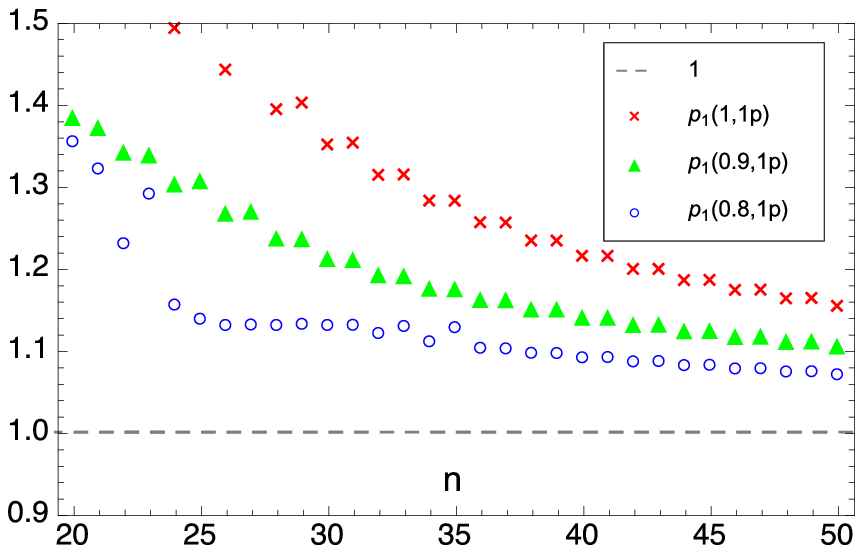}
\includegraphics[scale=0.8]{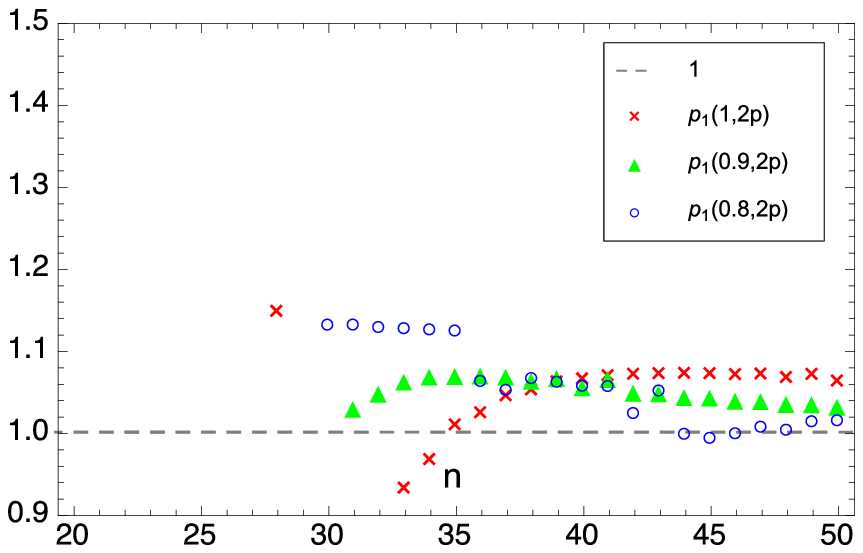}
\caption{Estimation results of $p_{1}=1$ for $r=1,0.9,0.8$ at $1$- and $2$-parameter ansatze (respectively plotted in the upper and lower).  Here $p_{1}(r,kp)$ denotes the estimate at Wilson parameter $r$ with $k$-parameter ansatz.}
\end{figure}

To summarize, we can say as follows:   As $r$ gets smaller, the region of continuum scaling observable in $\bar\beta_{n>}$ becomes wider but $\bar\beta_{n>}$ and its derivatives  begin to show oscillation at a bit smaller value of $r=1$.  On the other hand, when $r$ is larger ($r>1$), the effective region of $\bar\beta_{n>}$ gets narrower and the scaling behavior becomes vague.  We found that $r\in(0.8,1.0)$ provides us a good estimation up to $50$th order.  As in the cases frequently met with the Ising models, the estimation of the dynamical mass is better than the estimation of the exponent $p_{1}$.

\section{Concluding remarks}
We first found that the cancellation of the linear divergence in $\bar\beta_{n>}$ remains effective also under the $\delta$ expansion.  As a consequence, the true logarithmic behavior of bare coupling has been observed in $\bar\beta_{n>}$.  We remark that the confirmation is explicit for the range of Wilson parameter $r\in (0.8,1)=I$.   It is yet unclear whether other values of $r$ is essentially useless even when the order is large enough.   Present $50$th order study tells us, however, other values are not effective for   practical use.  In the range $I$, the estimation of the dynamical mass $m_{D}$ is carried out in the 1- and 2-parameter ansatze and all the sequences indicate the convergence to the exact value.  It is interesting to note that, from $35$th to $36$th orders and $43$rd to $44$th orders, rather big changes happen for $r=0.8$ in $2$-parameter ansatz.

It would be interesting to examine the estimation with the use of exact value of $p_{i}$ $(i=1,2)$.  In this case, we employ PMS in a looser variation of (\ref{first_cond}), $[1+p_{1}^{-1}\frac{d}{d\log t}]\bar\beta_{n>}^{(1)}+\frac{1}{2\pi}\sim 0$ for 1-parameter ansatz and $[1+p_{1}^{-1}\frac{d}{d\log t}][1+p_{2}^{-1}\frac{d}{d\log t}]\bar\beta_{n>}^{(1)}+\frac{1}{2\pi}\sim 0$ for the 2-parameter ansatz.  The result for $r=1$ is shown in FIG. 5 together with the result in previous full PMS protocol.  In the 1-parameter ansatz, the two sequences prove almost same accuracy (Present protocol with $p_{1}=1$ fixed gives slightly better result).  In the 2-parameter ansatz, the result with the exact input $p_{1}=1$ and $p_{2}=2$ shows better behavior to the orders of $20$th or so.  However, the two sequences tend to similar behaviors at larger orders.  
\begin{figure}
\centering
\includegraphics[scale=0.8]{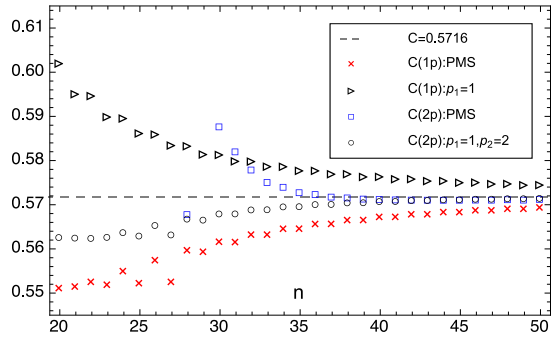}
\caption{Plots of estimation results for $r=1$ in the 1-parameter ansatz with the input $p_{1}=1$ and 2-parameter ansatz with the input $p_{1}=1$ and $p_{2}=2$.  For comparison, the results in 1- and 2-parameter ansatz in the previous section with the full PMS protocol are also plotted.}
\end{figure}
As this examination manifests itself and from the results so far obtained, the accuracy in the Gross-Neveu model with Wilson fermion is not good as in the Ising models.  Actually from numerical tests, this is roughly understood from the behaviors of $\bar\beta_{n>}^{(k)}$ $(k=1,2,\cdots)$ in such a way that the scaling behavior is not so clear to a few tens of orders.   The reason behind would be that non-oscillation of relevant functions needs $r$ around the value $r\sim 1$ and in the region the lattice artifact remains effective.



\begin{thebibliography}{99}
\bibitem{comment}  Our error in ref. \cite{yam2} has come from naive computational mistake.  We have also found that the constant $C_{2}$ appeared in eq. (62) has a typographical error and should be corrected as $C_{2}=\int_{-\pi}^{\pi}\frac{d^2 p}{(2\pi)^2}\Big[\frac{r\sum_{\mu}(1-\cos p_{\mu})}{(r\sum_{\mu}(1-\cos p_{\mu}))^2+\sum_{\mu}\sin^2 p_{\mu}}\Big]^2$.  In fact, the constant $C_{2}$ in ref. \cite{yam2} is nothing but $c_{2}$ in the present paper.  All results in ref. \cite{yam2} are, however, correct.
\end{thebibliography}
\end{document}